\documentclass[]{spie}  %

\usepackage{amsmath,amsfonts,amssymb}
\usepackage{graphicx}
\usepackage[colorlinks=true, allcolors=blue]{hyperref}

\newcommand{\ocs}{\texttt{ocs}}
\newcommand{\socs}{\texttt{socs}}

\title{The Simons Observatory: Deployment of the observatory control system and supporting infrastructure}

\author[a]{Brian J. Koopman}
\author[a]{Sanah Bhimani}
\author[e,f]{Nicholas Galitzki}
\author[d]{Matthew Hasselfield}
\author[a]{Jack Lashner}
\author[c]{Hironobu Nakata}
\author[a]{Laura Newburgh}
\author[a]{David V. Nguyen}
\author[b]{Tai Sakuma}
\author[b,g]{Kyohei Yamada}

\affil[a]{Wright Laboratory, Department of Physics, Yale University, New Haven, Connecticut 06511, USA}
\affil[b]{Joseph Henry Laboratories of Physics, Princeton University. Princeton, NJ 08544, USA}
\affil[c]{Department of Physics, Faculty of Science, Kyoto University, Kyoto 606-8502, Japan}
\affil[d]{Center for Computational Astrophysics, Flatiron Institute, 162 5th Ave 9th floor, New York, NY 10010, USA}
\affil[e]{Department of Physics, University of Texas at Austin, Austin, TX, 78712, USA}
\affil[f]{Weinberg Institute for Theoretical Physics, Texas Center for Cosmology and Astroparticle Physics, Austin, TX 78712, USA}
\affil[g]{Department of Physics, The University of Tokyo, Tokyo 113-0033, Japan}

\authorinfo{Corresponding author: \href{mailto:brian.koopman@yale.edu}{brian.koopman@yale.edu}}

\usepackage{aas_macros}

\begin{document} 
\maketitle

\begin{abstract}
The Simons Observatory (SO) is a cosmic microwave background (CMB) observatory
consisting of three small aperture telescopes and one large aperture telescope.
SO is located in the Atacama Desert in Chile at an elevation of 5180m.
Distributed among the four telescopes are over 60,000 transition-edge sensor
(TES) bolometers across six spectral bands centered between 27 and 280 GHz.
A large collection of ancillary hardware devices which produce lower rate
``housekeeping'' data are used to support the detector data collection.

We developed a distributed control system, which we call the observatory
control system (\ocs{}), to coordinate data collection among all systems within
the observatory. \ocs{} is a core component of the deployed site software,
interfacing with all on-site hardware. Alongside \ocs{} we utilize a combination of
internally and externally developed open source projects to enable remote
monitoring, data management, observation coordination, and data processing.

Deployment of a majority of the software is done using Docker containers. The
deployment of software packages is partially done via automated Ansible
scripts, utilizing a GitOps based approach for updating infrastructure on site.
We describe an overview of the software and computing systems deployed within
SO, including how those systems are deployed and interact with each other. We
also discuss the timing distribution system and its configuration as well as
lessons learned during the deployment process and where we plan to make future
improvements.
\end{abstract}


\section{INTRODUCTION}
\label{sec:intro}

The Simons Observatory (SO) is a modern cosmic microwave background (CMB)
observatory located in the Atacama Desert in Chile. The observatory consists
of three small aperture telescopes (SATs)\cite{galitzki2024simons} and one
large aperture telescope (LAT)\cite{2021RNAAS...5..100X,Zhu_2021}, all located
at an elevation of 5180m on Cerro Toco. This combination of small and large
aperture telescopes enables a wide range of science goals including measuring
the primordial perturbations, measuring the number of relativistic species and
the mass of neutrinos, improving constraints on parameters such as $H_0$, and
more \cite{1808.07445, 1907.08284, hensley_et_al_2022}. The four
telescopes combined house approximately 60,000 transition edge sensor (TES)
bolometers spread across six spectral bands from 27 to 280 GHz
\cite{2024JLTP..tmp...97D, 2022JLTP..209..815H, Dutcher_2023}. Currently, two
of the three SATs are actively observing, with the third coming online soon.
The LAT is expected to begin observations in early 2025.

Supporting the operation of the observatory is a complex network of hundreds of
devices. This diverse collection of hardware includes complex critical systems
such as the microwave multiplexing SMuRF readout
systems\cite{smurf,McCarrick_2021,Kernasovskiy_2018}, the telescope control
system\cite{acu}, and cryogenic rotating half-wave plate support
systems\cite{hwp,2024JLTP..214..173S}, as well as simple devices such as the
cooling loop flowmeters. We designed, and have now deployed, a distributed
observatory control system, called the observatory control system (\ocs{}), to
monitor and control this wide range of devices \cite{2020SPIE11452E..08K}.

In this proceedings we present an update on the state of \ocs{} since Koopman
et al. 2020\cite{2020SPIE11452E..08K} in Section \ref{sec:ocs}, describing
progress made in expanding its functionality over the past four years. In
Section \ref{sec:deployment} we describe the nearly full scale deployment of
\ocs{} on site at SO, including how we deploy the agent configuration files.
Then, in Section \ref{sec:services} we discuss the various services that
support observatory operations. We also discuss lessons learned throughout
deployment in Section \ref{sec:lessons} and plans for future work in Section
\ref{sec:future}, before concluding in Section \ref{sec:summary}.

\section{OCS UPDATES}
\label{sec:ocs}
The observatory control system (\ocs{}) is a distributed control system designed
and built for use on the Simons Observatory. It consists of a set of
``agents'', which are long-running processes that implement functions that can
be called and monitored remotely. Many agents use driver code to provide
remote access to a given hardware device. The agents all connect to a
middleware layer, a Web Application Messaging Protocol (WAMP) router provided
by a crossbar server, that enables remote procedure call and message passing
via PubSub. Control programs then connect to this WAMP router to orchestrate
observatory operations by commanding the network of agents. \ocs{} has been
used extensively in the labs developing the SO telescopes and is now deployed
at nearly full scale at the SO site. In this section we describe the updates to
\ocs{} since Koopman et al. 2020\cite{2020SPIE11452E..08K}.

\subsection{Agents}

Agents are one of the core components of \ocs{}. Each agent uses either
internally developed or externally available driver code to interface with a
given hardware device. Agents can be a thin wrapper around the drivers or
provide more complex functionality when necessary. The agent exposes an API
that can then be used by control programs to run operations within the agent.
We previously reported on 17 agents, and have since added 32 new agents related
to hardware deployed on site. Total there are now 50 agents, with 8 agents
related to the HWP, 4 agents related to the wiregrid, 6 agents related to the
SMuRF systems, 10 agents related to supporting the cryogenics, and 22 other
agents. The full list of agents and the functionality they provide is shown in
Appendix \ref{sec:appendix_a}.

Where possible we chose devices that provide a network interface. Such devices
typically have a simple TCP or HTTP communication interface. This allows the
corresponding agent to run anywhere on the network, rather than requiring it to
run on a computer physically close to the device.  Not all hardware on site
provides a network interface, so we still have a collection of devices with
serial interfaces (either RS-232 or USB). In these instances, we typically run
a small Intel NUC or single board computer such as a Raspberry Pi or BeagleBone
close to the device. The agent associated with the hardware then runs on this
nearby computer.

There are still a number of Agents in development, which we expect to be added
in the coming months as we transition into normal operations mode. All agents
are available in either the
ocs\footnote{\url{https://github.com/simonsobs/ocs}} or
socs\footnote{\url{https://github.com/simonsobs/socs}} repositories.

\subsection{New Features}

There are several new features in the core library and in core \ocs{} agents
that have been introduced over the past four years. In this section we describe
some of those updates.

In the core library a new plugin system was introduced. This system enables the
creation of third party Python packages that contain additional agents. Package
metadata is used to announce that a package is an \texttt{ocs} plugin. The
agents provided by the package are then made available to launch with the new
`\texttt{ocs-agent-cli} tool, the new recommended mechanism for launching
agents.  \socs{} is an example of such a plugin package. Another new
command-line interface (CLI) tool, \texttt{ocs-client-cli}, was also
introduced. This provides a simple way to initialize an \texttt{OCSClient}
object, list all online agents, and to subscribe and view any data feed. 

Within the core library, support for passing booleans on data feeds has been
added. The ``address root'', a part of the configuration meant to define the
start of an agent's address on the crossbar server, is now configurable, where
previously it was fixed to be ``observatory''. This, when used, prevents
collisions when viewing data from \ocs{} in Grafana, where the measurement
names are a combination of address root and agent instance-id.

Other core library changes include the addition of a decorator that can be
added to operations that checks the validity of parameters passed to the
operation. This prevents invalid data types or out of range values, which would
typically require explicit error handling for these cases, from being passed to
an agent at all, instead causing an error on the client end. If an agent loses
its connection to the crossbar server by default it tries to reconnect for ten
seconds before shutting down. This timeout period is now configurable and can
be disabled. If disabled, the agent will stay online even without the crossbar
server, allowing agents like the ACU agent to complete operations such as
telescope motion, even if crossbar is unavailable.

There are a few new features in core agents that are worth mentioning.  The
Host Manager Agent, renamed from ``Host Master Agent'', has been overhauled
significantly to include support for running Agents within Docker containers,
as has become standard for all agents deployed at site. This includes support for
Docker Compose v2. The InfluxDB Publisher Agent was updated to support Influx's
`line' protocol, increasing writing efficiency to the database. OCS feeds can
also now be excluded explicitly from being written to InfluxDB when
instantiated. This allows high sample rate data to still be sent to the
Aggregator agent while a downsampled rate is sent to InfluxDB, which struggles
with higher sample rates ($>10\,\mathrm{Hz}$ or so.) Lastly, the Registry agent
now tracks the status of each operation in every agent on the network. This
includes a new ``degraded'' status that is used to indicate a non-fatal problem
with operations that have halted data collection.

\subsubsection{ocs-web}
ocs-web, the web interface for \ocs{}, has been replaced with a new Vue.js
based web application\footnote{\url{https://github.com/simonsobs/ocs-web}}. The
new version supports easily switching between multiple different crossbar
servers. This feature allows one ocs-web instance to support all of SO, where
we run five crossbar servers (see details in Section \ref{sec:deployment}.)

\begin{figure}[!htbp]
\begin{center}
\begin{tabular}{c} %
\frame{\includegraphics[width=0.95\linewidth]{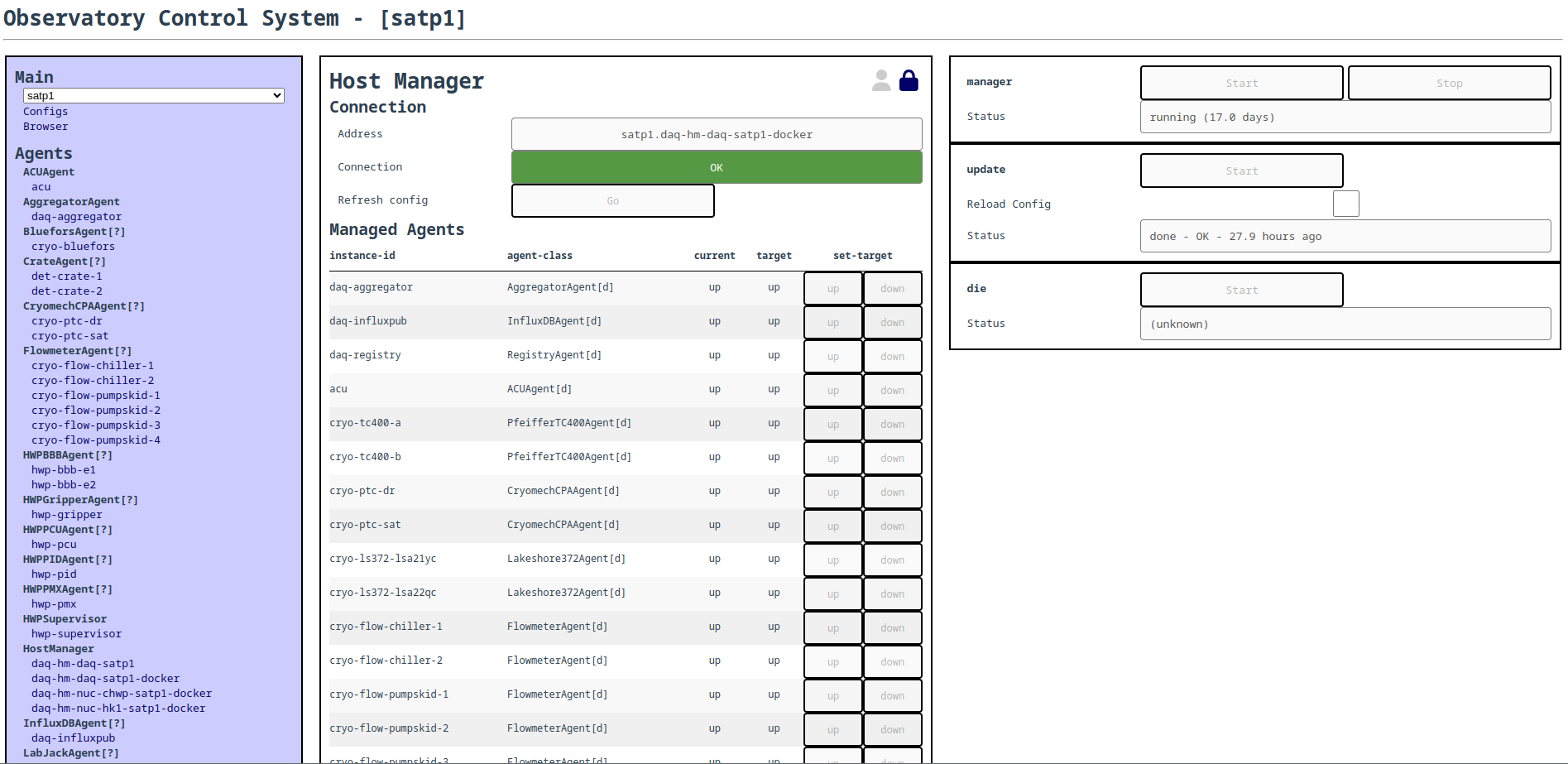}}\\
\frame{\includegraphics[width=0.95\linewidth]{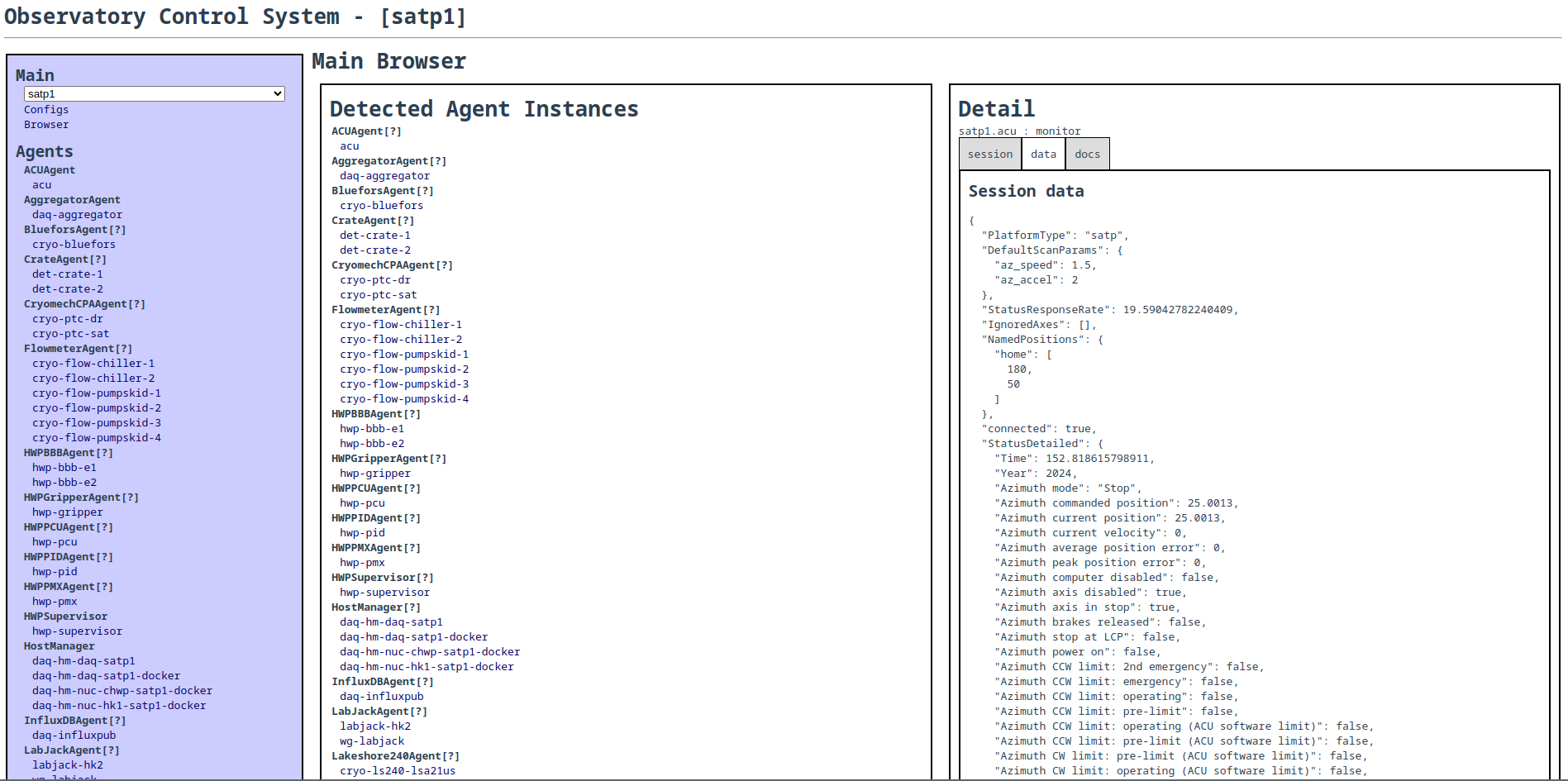}}
\end{tabular}
\end{center}
\caption[example]
{ \label{fig:ocs-web-1}
Screenshots of ocs-web with the panel for the main Host Manager agent on SATP1
(small aperture telescope platform one, the general name we use for referring
to the first SAT) visible (top) and the main browser (bottom). The left panel
shows all active agents. The center panel shows all agents managed by the Host
Manager (top) or a list of all agents (bottom). In the top screenshot, users
can unlock the panel by clicking the lock at the top of the panel, then they
may bring any agent down or up. This is useful for restarting agents when
needed. The right panel shows the available Host Manager operations, which can
also be run from this page. In the bottom screenshot, clicking on an agent
brings up a list of the agent's operations, which can be selected to show the
session message buffer, session data, and documentation for the operation. This
particular screenshot shows the session data from the SATP1 ACU agent.
}
\end{figure}

Basic functionality remains relatively unchanged since Koopman et al.
2020\cite{2020SPIE11452E..08K}. Once connected to the crossbar server a list of
running agents is shown on the left side of the screen. Clicking any one of the
agents brings up a panel for the agent, either a generic panel or a panel
designed for the unique properties of the agent. The panel displays relevant
information about the agent and allows users to run any of the operations
available in the agent. Screenshots are shown in Figure \ref{fig:ocs-web-1}.

\subsubsection{Testing}
\label{sec:testing}
We provide some infrastructure in the core \ocs{} library to help with testing
agents. This includes pytest fixtures for running individual agents as a
subprocess during integration tests as well as a fixture for instantiating a
client to command the running agent. We also include a fixture for checking the
connection to a crossbar server, this can be used to ensure a crossbar server
is online before integration tests are run.

The \socs{} library also contains a fixture that creates a ``device emulator''
for use in integration tests.  Typically testing an agent requires access to
the hardware it communicates with. This can be difficult, as there is often
limited access to this hardware -- either it is in use during
testing in the labs or during observations in the field. Unit tests could be
used to test individual aspects of the agent code, mocking out the
communication layer, but this does not catch certain parts of the agent, such
as passing parameters from a client to the agent, which if done incorrectly can
prevent an operation from running.  To catch this more complex behavior
integration tests are required, but that requires a crossbar server instance,
the agent, and either access to the hardware or some mock interface. Since
hardware access is difficult, we developed the device emulator fixture to allow
a developer to mock the responses expected from the device.

This pytest fixture starts up a serial or TCP relay and allows the developer to
define a set of messages and expected responses to those messages. The
integration tests then start up a crossbar instance, the device emulator, and
then the agent. The crossbar server persists per pytest session, but the device
emulator is restarted for each individual test to avoid tests interfering with
each other.

\section{SITE DEPLOYMENT}
\label{sec:deployment}

Over the past year we have deployed a majority of the observatory control
system for SO. The network of agents required to run the entire observatory is
necessarily much larger than those used in any individual lab during
development. While we did perform some scale testing, this was the first time
we deployed all of the agents at scale. In this Section we discuss the layout
of the control system across the entire observatory, and then discuss the
configuration deployment.

\subsection{Site Agent Layout}

\begin{figure}[!htbp]
\begin{center}
\begin{tabular}{c} %
\includegraphics[width=0.98\linewidth]{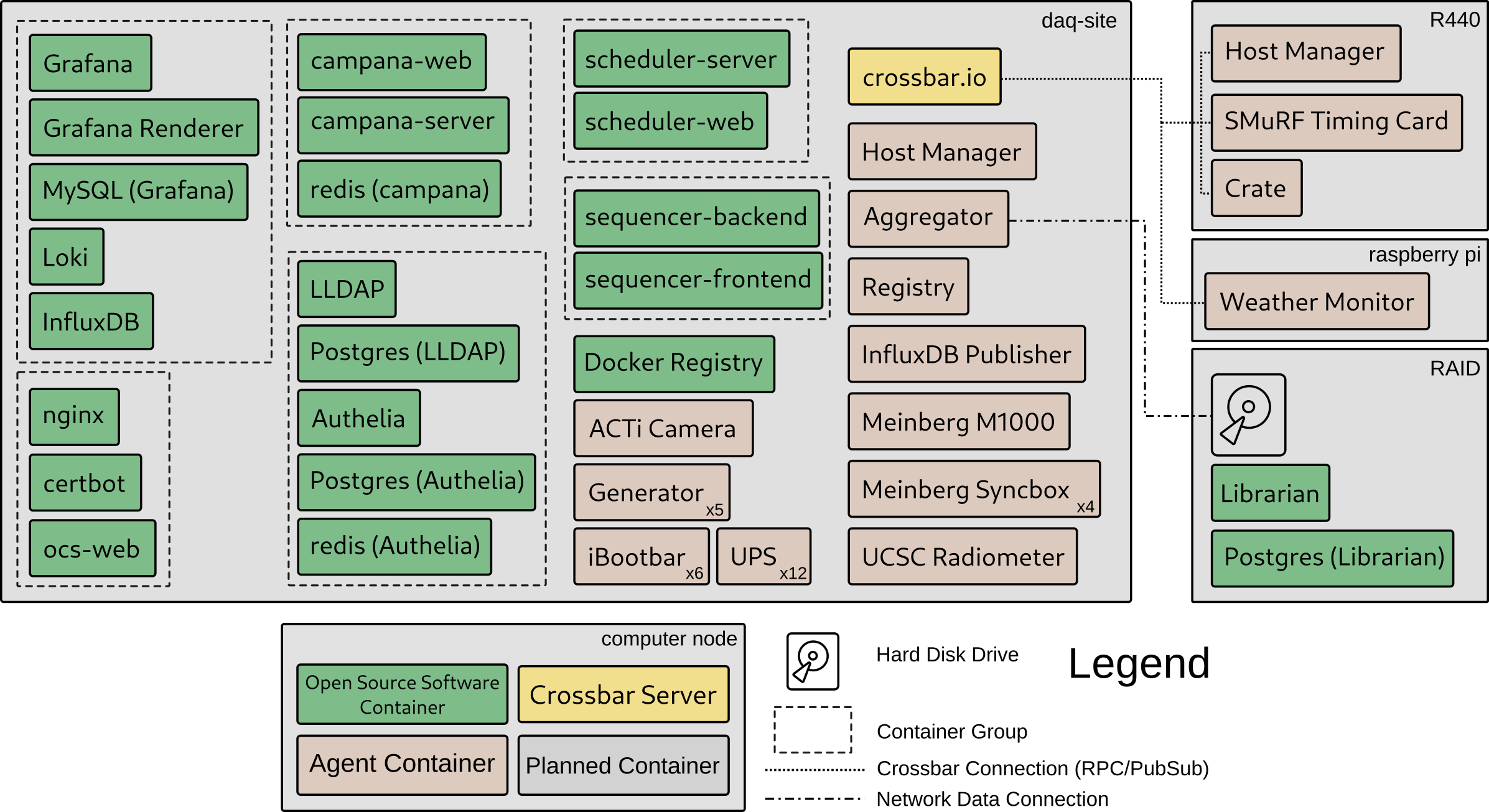}
\end{tabular}
\end{center}
\caption[example]
{ \label{fig:site-shared}
A diagram showing the `site' \ocs{} and supporting services Docker container
distribution. Most services are run on a single virtual machine, labeled
`daq-site'. This includes groups of containers supporting Grafana for
monitoring, Authelia for authentication, the scheduler for observation
scheduling, the sequencer for executing schedules, and nginx for proxying the
various web services. Also shown here are the \ocs{} containers running for
site wide hardware, such as the Meinberg timing systems, as well as the RAID
array, which runs the Librarian software.}
\end{figure}

\begin{figure}[!htbp]
\begin{center}
\begin{tabular}{c} %
\includegraphics[width=0.80\linewidth]{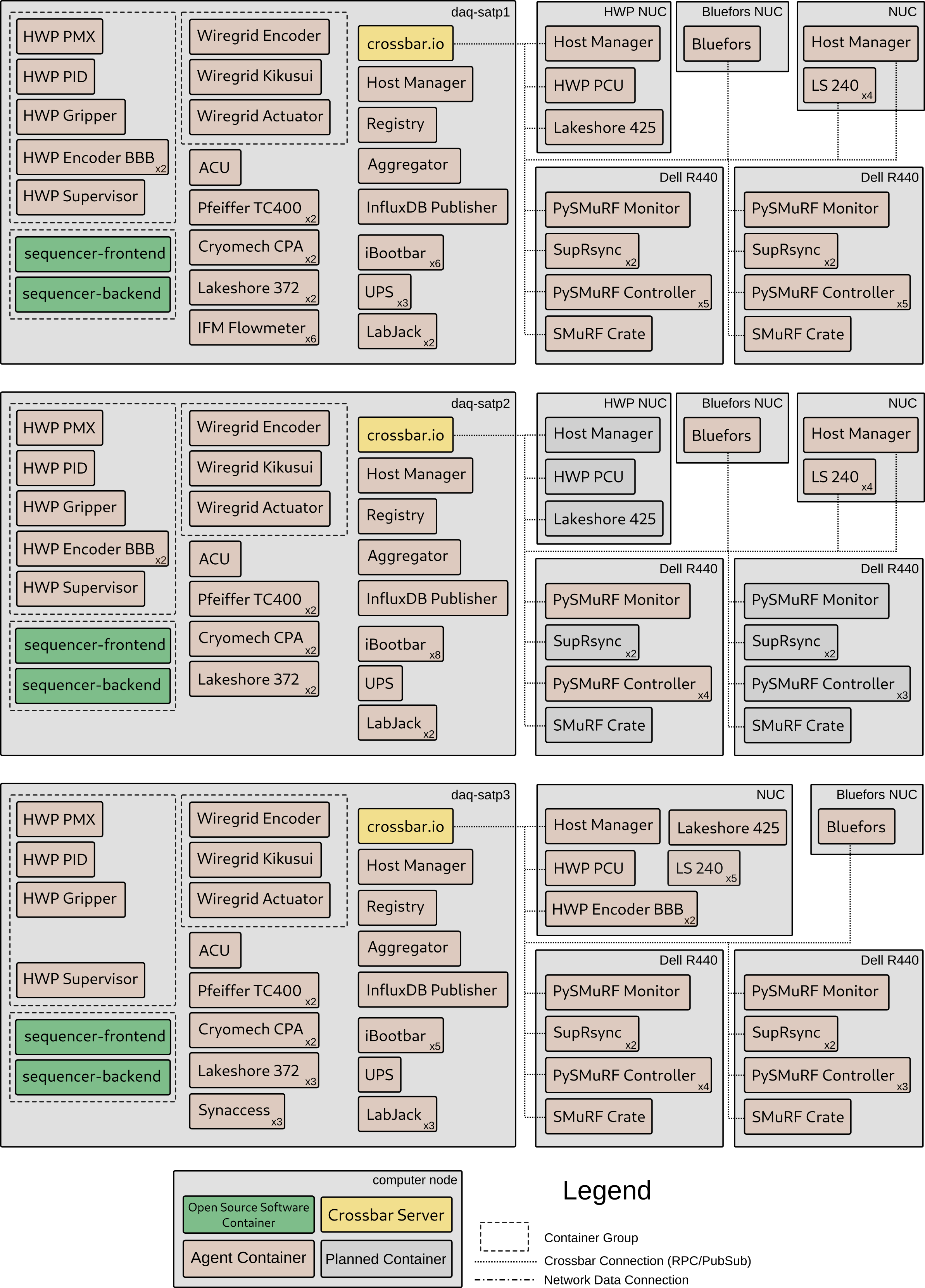}
\end{tabular}
\end{center}
\caption[example]
{ \label{fig:ocs-sats}
A diagram showing the \ocs{} network for all three SATs. Each SAT has a core
virtual machine that runs the crossbar server for a given SAT, as well as the
sequencer for the platform and most of the agents. Additionally each SAT has a
collection of NUCs to support HWP readout, Lakeshore 240 low temperature
thermometer readout, and monitoring of the Bluefors dilution refrigerator logs.
Lastly, there are two Dell R440 servers that support the detector readout
through the SMuRF systems, which we refer to as SMuRF servers. Each SMuRF
server runs a collection of PySMuRF controller agents that perform detector
operations, such as IV curves and data streaming, two SupRsync agents and a
PySMuRF monitor agent to move data off of the local disk to the RAID, and a
crate monitor agent.}
\end{figure}

\begin{figure}[!htbp]
\begin{center}
\begin{tabular}{c} %
\includegraphics[width=0.98\linewidth]{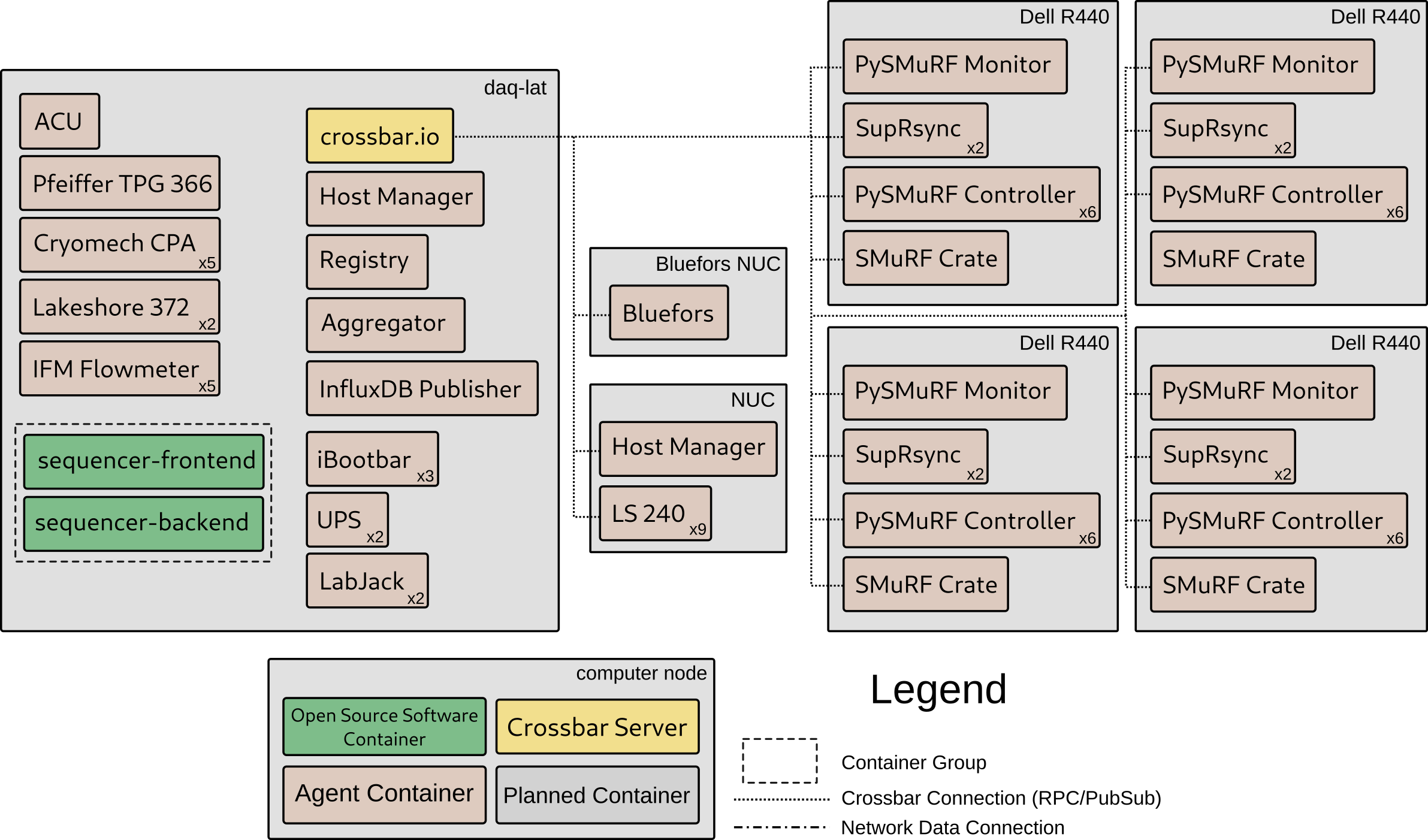}
\end{tabular}
\end{center}
\caption[example]
{ \label{fig:ocs-lat}
A diagram showing the \ocs{} network for the LAT. The layout is similar to any
single SAT layout shown in Figure \ref{fig:ocs-sats}. The main differences
being there are no HWP or wiregrid agents and there are more SMuRF servers to
support the large number of detectors in the LAT.}
\end{figure}

The \ocs{} agent layout at the site spans at least 28 computers. Each platform
has a primary machine that runs as a KVM guest on one of two large Linux
servers. These primary machines each run a crossbar server, the middleware
layer for \ocs{}, along with most of the agents that connect to their
corresponding devices over the network. Each SAT platform has two Dell R440
servers to handle control and data acquisition for the detector readout system,
which we refer to as ``SMuRF servers'', while the LAT has four. There is an
additional SMuRF server shared among the site for timing distribution. Agents
that require special connections to hardware, often requiring close proximity
to that hardware, are run on small Intel NUC or single board computers nearby
or on the platforms.  Across the entire observatory there are currently over
250 configured agents.  The layout of these agents is shown in Figures
\ref{fig:site-shared}, \ref{fig:ocs-sats}, and \ref{fig:ocs-lat}. For a
detailed look at the SATP1 agent deployment see Bhimani et al.
2024\cite{bhimani_et_al}.

Figure \ref{fig:site-shared} shows the site wide agents, such as the Meinberg
M1000 and Syncbox agents, the radiometer agent, ACTi Camera agent, diesel
generator agents, weather monitor, and SMuRF timing agents. Additionally it
shows the large collection of open source supporting services that are run to
facilitate remote operations. These are detailed in Section \ref{sec:services}.

Figure \ref{fig:ocs-sats} shows the layout of agents for the three SATs. Each
crossbar server and most of the agents run on a single KVM guest on one of two
main servers located in a central server room. The sequencer frontend and
backend are also served from these virtual machines. Each SAT has two SMuRF
servers, which run the required agents for interfacing with the detectors. 

There are small differences between the agents run for each platform, such as a
different distribution of HWP agents on SATP3. Other differences include the
use of several Synaccess agents in place of iBootbar agents. These differences
typically reflect small differences in the hardware configurations between
platforms, which were all assembled by different teams.

The LAT agent layout is shown in Figure \ref{fig:ocs-lat}. This is similar the
SAT layouts without the HWP or wiregrid agents, as the LAT contains neither of
those optical elements. There are also two additional SMuRF servers, required
to readout the increased number of detectors in the LAT compared to an
individual SAT.

\subsection{Configuration Deployment}
Management of a distributed control system presents an interesting system
administration problem. Configuration files are required on each computer
running \ocs{} agents. Users who need to edit the configuration files need
proper permissions to do so, and ideally a history of changes is kept so they
can be rolled back if a change adversely affects the system. We take a GitOps
based approach to deploying configuration changes.

We use a private GitHub Actions runner to execute an Ansible playbook on the
site computers to deploy configuration changes. This allows us to manage the
users who have permissions to change configuration files using GitHub's teams
feature in combination with permissions settings on the repository used to
store the configuration files. The user configured to run the playbook through
the Actions runner has write permissions to the configuration files on each
computer, allowing the playbook to be executed.

The workflow to deploy a change is to submit a pull request on the
configuration repository. Once the change is reviewed and approved the pull
request is merged. On merge the private GitHub Actions runner clones the
latest \texttt{main} branch and executes the Ansible playbook which copies the
updated files to the required hosts. Once the change is deployed, any agents
that have had their configuration changed will need to be restarted, which can
be done through ocs-web with the Host Manager agent. Depending on the agent,
this might require observations to be halted, for instance if the ACU agent
configuration changes. The update can often be done at the next convenient
time, i.e. during any downtime.

\subsection{Timing Distribution}
The Meinberg M1000 is the main source of timing for the entire SO site. The
OCXO-HQ oscillator within it is synchronized to GPS. The M1000 then outputs a
10 MHz signal which is required to discipline a 122.88 MHz clock in the SMuRF
timing system. Additionally the M1000 outputs a PTPv2 (IEEE 1588-2008, we will
just refer to this as PTP) signal on the network. This PTP signal is received
by the four Meinberg Syncbox/N2X signal converters on each platform and by the
antenna control units (ACUs). The Syncboxes discipline their internal
oscillator to this PTP signal, and have three configurable BNC outputs. These
outputs are configured as needed on platform. One example of a required output is
an IRIG-B signal required by the HWP subsystem.

PTP distribution on the network requires PTP compatible networking equipment.
All network switches within the PTP path are PlanetTech brand industrial
switches that are PTP compatible. The switches must be properly configured for
PTP passthrough, else the quality of the PTP timing signal past the switch may
be degraded from the $<\pm100\,\mathrm{ns}$ synchronization specification to
$\sim\pm1\,\mathrm{ms}$. An ansible playbook is used to configure the switches
via their Telnet interface. The final configurations were tested against a
secondary GPS clock temporarily located next to the Syncboxes, where a pulse
per second (PPS) signal could be compared between the two. With the switches
properly configured this difference was seen to be on order $100\,\mathrm{ns}$.

\section{SUPPORTING SERVICES}
\label{sec:services}

Along with OCS we employ the use of many different supporting services to
monitor and control the observatory. These services include tools to
orchestrate observations, alert when things go wrong, and securely access site
services remotely. In this section we discuss details about these services and
how they are deployed.

\subsection{Access and Authentication}
Secure remote access is required to run the observatory. Remote observers
around the world keep the telescopes running, ensuring that observation
schedules are carried out and resolving any potential issues that arise. An
\texttt{nginx} web server functions as a reverse proxy for the site's web
services. Along with \texttt{nginx} we use the open-source authentication and
authorization server Authelia\footnote{\url{https://www.authelia.com/}}.

Authelia supplements \texttt{nginx} by sitting between the reverse proxy and the
application that is being secured. Requests get forwarded from \texttt{nginx} to
Authelia to check for authentication. If the user is not logged in they will be
redirected to do so. Logging in requires multi-factor authentication in the
form of password and time based one time password. Once authenticated, future
requests pass the authentication check when forwarded to Authelia, at which
point the proxy forwards the request to the protected application.

Authelia also supports OAuth 2.0 and OpenID Connect 1.0, which allows it to
integrate into applications that support them, such as Grafana and JupyterHub.
Deploying an OAuth 2.0 solution locally on site has the advantage of still
being accessible even if the Internet connection to site is lost as compared to
using one of the many popular OAuth providers such as GitHub. Authelia is
backed by an LDAP server on site which stores usernames and credentials.

\subsection{Monitoring, Alerting, and Log Aggregation}
Our monitoring and log aggregation system remains largely unchanged since
Koopman et al 2020. We continue to use Grafana, backed by an InfluxDB, for
monitoring along with Loki for log aggregation. The integration between Grafana
and Loki has improved over the years as Grafana continues to release updates
for both software packages.

\setcounter{footnote}{0} 

The SMuRF systems output several ancillary files in addition to the
timestreams. These files are typically text files and plots, which are useful
to inspect to assess health of the system and to debug any problems that
arise. We developed a web portal for viewing these outputs called
TeleView\footnote{\url{https://github.com/simonsobs/TeleView}}. TeleView uses
the Django web framework to provide an interface to the MongoDB database, which
stores metadata about the SMuRF files. A Next.js front end serves as the user
interface. TeleView can ingest arbitrary files, but its main use is for the
SMuRF outputs.

We use the alerting functionality within Grafana, coupled with a custom built
application called Campana\footnote{\url{https://github.com/simonsobs/campana}}
for sending SMS and phone calls, to send alerts to remote observers when issues
arise with the observatory. This mechanism queries the housekeeping data stored
in the InfluxDB backend at regular intervals, evaluates rules for alerting, and
then sends a Slack message or triggers an alert in Campana. For more details on
how this system works see Nguyen et al.  2024 \cite{nguyen_et_al}.

\subsection{Data Movement and Processing}
A fork of the HERA Librarian is used to manage data collected on site. Final
data products, which we refer to as ``books'', are managed by the Librarian,
which tracks the books in a Postgres database. During commissioning, the
Librarian would then copy the data to a ``sneakernet'' disk, which would be
shipped or hand carried back to North America to the San Diego Supercomputer
Center and copied from there to NERSC.

A high speed fiber, providing 300 Mbps symmetric speeds to North America, was
recently installed. This fiber connects through the ALMA high elevation
facility and is provided in coordination with REUNA. This will enabled rapid
copying of the data via the Librarian to NERSC. For more details on the Librarian
see Borrow et al. 2024\cite{borrow_librarian}.

We deploy several separate JupyterHub installations, split across three KVM
guests running on two servers, to enable user interactive data exploration.
Remote users can access all on-site data via these JupyterHub instances, and
use them to load and analyze the data shortly after it is collected. There is
an ongoing effort to move to a single Kubernetes-based JupyterHub installation
that spans the now four ``compute nodes'' on site. This will ease
administration and user management in the system. A common problem encountered
in the current configuration is complete consumption of memory on a node, which
causes all Jupyter kernels on that instance to crash. The Kubernetes
installation should mitigate this in several ways, distributing users evenly
among the compute nodes, and providing guarantees on allocated memory per user.

These local compute resources, while still important for automated data
processing and local access, have seen decreased use after a recent connection
to high speed fiber was made, enabling movement of data at site to computing
facilities at NERSC, Princeton, and the University of Manchester much more
quickly than was previously possible.

We have adopted the Prefect workflow orchestration
system\footnote{\url{https://www.prefect.io/}} for managing automated data
processing and data packaging scripts. Prefect provides cronjob-like
functionality, with the ability to track dependencies between jobs, and to
retry failed jobs. The web interface allows remote observers to easily view the
health of the automatic data processing scripts, and Prefect can send alerts on
failure. For more details on how we use Prefect see Guan et al.
2024\cite{guan_et_al}.

\subsection{Observation Sequencing}
\label{sec:sequencer}

Many elements must work in concert to perform observations; the detectors must
be tuned, the telescope pointed, and ancillary hardware must be controlled.
\texttt{sorunlib} is a Python library that forms the basis for writing a
``schedule''. The schedules are written in Python \cite{guan_et_al} and are
serially executed scripts that perform each step required during an
observation.  \texttt{nextline} is the tool used to execute these scripts.
\texttt{nextline} has a web interface that users interact with to upload
schedules and to see the state of the current observation.

\subsubsection{sorunlib}
\texttt{sorunlib}\footnote{\url{https://github.com/simonsobs/sorunlib}} is a
high level Python library used to run observations. The library is responsible
for creating the necessary ocs clients for any hardware that needs to be
coordinated during an observation. It makes this list available as a global
\texttt{CLIENTS} variable. The hardware subsystems that require coordination
each have their own module within \texttt{sorunlib}, which access this global
clients dictionary and command the relevant ocs agents.

\texttt{sorunlib} checks the response of each command for the operation status
and opcode (operation code), handles any errors that occur, and depending on
the severity of the error, raises an exception within the schedule, halting
observations. There are modules for commanding the ACU, HWP, SMuRF, and wiregrid,
as well as a module for sequencing detector operations with telescope motion.
The goal of \texttt{sorunlib} is to provide a high level API that presents an
easy to read and concise ``schedule'' that remote observers can follow along
with to understand what is happening during observations.

\subsubsection{nextline}
\texttt{nextline} is a custom written tool for executing Python scripts
line-by-line. The \texttt{nextline} Python package is the core for the
\texttt{nextline-graphql} backend server. This server provides an API for the
\texttt{nextline-web} frontend, written in TypeScript, to interact with.
Schedules are uploaded from the web frontend, executed within the nextline
backend, and their status displayed for the user to see back on the frontend
\cite{nextline}.

\texttt{nextline} uses an asynchronous wrapper for pytest's \texttt{pluggy}
system, called
\texttt{apluggy}\footnote{\url{https://github.com/simonsobs/apluggy}} to extend
functionality with various plugins. Table \ref{tab:nextline-plugins} shows a
list of external plugins and their functionality.

\begin{table}[ht]
\caption{List of external \texttt{nextline} plugins.}
\label{tab:nextline-plugins}
\begin{center}
\begin{tabular}{|l|l|}
\hline
\textbf{Plugin} & \textbf{Description} \\
\hline
\href{https://github.com/simonsobs/nextline-rdb}{\texttt{nextline-rdb}} & Relational database for configuration, execution history, and other information.\\
\hline
\href{https://github.com/simonsobs/nextline-schedule}{\texttt{nextline-schedule}} & Fetch pre-made schedules from the scheduler API.\\
\hline
\href{https://github.com/simonsobs/nextline-alert}{\texttt{nextline-alert}} & Send alert triggers to \texttt{campana} when an exception is raised.\\
\hline
\end{tabular}
\end{center}
\end{table}

\subsubsection{Observation Scheduling}
Observatory operations are executed by the sequencer, described in Section
\ref{sec:sequencer}. The ``schedules'' run by the sequencer are generated by
the ``scheduler''. The scheduler is primarily a Python library that takes a
master observation plan, a list of what targets to observe when, and translates
them it into an sorunlib based schedule that can be run by the sequencer. A
simple Flask API presents an interface for the scheduler web front end to fetch
schedules. The nextline web front end also pulls schedules directly from this
API. For more details on the scheduler, see Guan et al. 2024\cite{guan_et_al}.

\subsubsection{Deploying the Sequencer}

The combination of \texttt{nextline} and \texttt{sorunlib} is deployed in two
Docker containers, one for the frontend and one for the backend. The frontend
container is the one published in the GitHub container registry for the
\texttt{nextline-web} package. The backend container runs from a custom built
image, based on the one provided by the \texttt{nextline-graphql} repository,
that installs \texttt{sorunlib} and any required external plugins. This is
built and released from a separate repository called
\texttt{so-daq-sequencer-docker}\footnote{\url{https://github.com/simonsobs/so-daq-sequencer-docker}}.

\section{LESSONS LEARNED}
\label{sec:lessons}

We have been supporting \ocs{} for over six years now. \ocs{} has been deployed
in some form at the site for just over a year. In that time we have learned a
few lessons that we share in this section to help those who may be developing
their own control systems, or those who may be considering use of \ocs{}.

\subsection{Development}
\label{sec:lessons-development}
There are over thirty contributors to \ocs{} and \socs{} combined. While we
have a style guide, many early reviews contained many style-based comments.
We have since adopted the use of pre-commit, a tool that manages git pre-commit
hooks. This is configured to run several simple checks such as checking if the
committed files are valid Python and removing trailing whitespace at the end of
lines. We have also configured it to run a couple of auto-formatters, namely
isort and autopep8, which sort imports and format code to conform to Python
PEP 8. flake8 then catches anything that autopep8 did not. We use the
integration of pre-commit with GitHub actions that runs pre-commit on each PR,
and require that it pass before merging. This has greatly reduced the need for
style-related comments in our reviews.

The testing infrastructure described in Sec. \ref{sec:testing} has been a
valuable way to test agent code without access to hardware, however it also
increases the time it takes to run the GitHub Actions during PRs, as we run all
unit and integration tests, following continuous integration (CI) best
practices. Improperly written communication code can also lead to difficult to
debug problems with the tests. Agent contributors tend to be graduate students
and postdocs working in the lab with the relevant hardware, but test
development has typically been done by dedicated members of the software team.
The contribution of agents from across the collaboration has really enabled the
larger number of agents supported by \ocs{}, but less experienced developers
might not be familiar with test development. As a result, our coverage remains
relatively low at $\sim35\%$ of the \socs{} agents. Had we developed this
testing capability earlier and provided thorough examples, it might have been
easier to obtain wider adoption by contributors.

\subsection{Deployment}
Uniformity in the method of configuration file deployment was a challenge that
is still not solved in our deployment. The core ``DAQ'' nodes were setup by the
team developing \ocs{}, and as such are the ones that are directly managed by
the automated deployment scheme described in \ref{sec:deployment}. However,
each platform team provided their own, often already configured, computers for
specific hardware related to their platform, e.g. the HWP NUCs, the SMuRF
servers. These machines vary in the details of their configuration, details
such as the user running OCS, the storage path of the configuration files,
which version of Ubuntu is running, and which version of Python is installed.
The configuration files on these machines are still managed by the platform
teams in whatever configuration they were in when initially set up. Making the
configuration deployment uniform across platforms is a work in progress, but
would have been more easily accomplished if considered from the start. This
experience also motivates developing a centralized configuration file system,
as mentioned in Section \ref{sec:middleware}.

\section{FUTURE WORK}
\label{sec:future}

\ocs{} is almost fully deployed at the SO site. We have approximately a decade
of observations ahead, through which we will continue to support \ocs{}. This
support includes typical maintenance, such as updating \ocs{} when new Python
versions are released and when dependencies require downstream modifications.
In this Section we discussion some additional considerations for the future of
\ocs{}.

\subsection{Middleware Replacement}
\label{sec:middleware}
Crossbar has been the selected middleware layer used by \ocs{}. Selected in
early 2018, crossbar was chosen due to its unique capabilities, providing both
remote procedure call (RPC) and PubSub. It was an open source implementation of
a WAMP router, written by the authors behind WAMP. Additionally, it was
accompanied by the Autobahn libraries, which provided support for multiple
programming languages to interface with the crossbar server.

However, in recent years the company developing crossbar has, as far as we can
tell,
dissolved\footnote{\url{https://github.com/crossbario/crossbar/issues/2085\#issuecomment-1676372430}}.
The project still sees minor updates, and remains available on GitHub, but the
main homepage for the project, \url{crossbar.io}, is now offline, so
documentation is inaccessible. Use of the crossbar server, deployed in an
isolated Docker container, has been remarkably stable over the six years we
have used it for SO. Unfortunately, these recent events will force us to
replace crossbar as a middleware layer.

We are still considering the best path forward, however, we expect to build a
replacement for this core RPC and PubSub functionality that crossbar provided
that can be dropped in as a replacement for crossbar. Examining recent trends
in the design of observatory control systems, it is difficult to identify the
`best' middleware layer. Technologies such as ZeroMQ and ICE are high on our
list of software to evaluate, and that we might expect to incorporate into our
crossbar replacement \cite{2021AdAst2021E..22L}.

Building our own solution allows us to more easily extend functionality as
well. Centralizing the configuration information would make deployment of the
agents simpler if it can be done in a manner that maintains the history of
configuration changes and has proper user authentication built into it. Of
course additional features come at the cost of additional developer time, which
is limited.

\subsection{Access Control}
Currently any client on the network that can connect to the crossbar server can
send commands to any agents. In labs, where teams using \texttt{ocs} are
generally small and well coordinated, this has not presented an issue. However,
at the full scale of the observatory, this presents potential issues of
multiple users sending redundant or conflicting commands. Remote observations
are carried out by a single responsible operator and so conflicts have yet to
arise in practice, but we do have plans to address this potential pitfall.

We are working on an access control system, designed to restrict control
programs from sending commands to certain agents at certain times. The system
is being built to prevent accidental mistakes, not prevent malicious behavior
on behalf of the user. Agents will define several levels of access, requiring
different credentials. Clients will need to pass the proper credentials to
override the current restriction level in order to send commands.
Implementation details are still being worked out.

\subsection{Kubernetes}
Kubernetes has grown in popularity since \ocs{} was originally developed,
seeing some use in observatory deployments\cite{2020SPIE11445E..0IT}. We
already deploy agents within Docker containers, but never made the leap to
managing the containers with Kubernetes. Maintaining a local Kubernetes cluster
is a large undertaking, and until the recent SO deployment to Chile, would
require each lab testing various components to run their own cluster, which was
not feasible. Now that we are in the field, and actively working on a
Kubernetes cluster for other purposes, the use of Kubernetes to manage some
aspects of \ocs{} becomes more appealing.

The Host Manager agent currently orchestrates the agent containers, and has the
added feature of easy user interaction to start and stop agents. We may never
move away from this, but running the middleware layer on a Kubernetes cluster
would be appealing to ensure high availability of this critical component. This
deployment strategy will be kept in mind as we work on the previously described
middleware replacement.

\section{SUMMARY}
\label{sec:summary}

The Simons Observatory has begun observations with two of the three SATs, with
the third coming online imminently. The LAT is expected to begin observations
in early 2025. We have presented an update on the status and deployment of the
observatory control system, including discussion of newly developed agents, new
features in the core library, and the state of supporting services for
observatory operations.

\ocs{} is now successfully deployed at nearly full scale and observations are
underway. While we do have future plans to replace the middleware layer and
plan to continue to add agents where needed to support additional hardware, we
consider the deployment a success. We will continue to support \ocs{}
throughout the life of SO, approximately the next ten years. We encourage those
looking for a control system to consider using \ocs{} and contributing to its
development.

\appendix

\section{Agent List}
\label{sec:appendix_a}

This list shows all current \texttt{ocs} and \texttt{socs} agents and describes their functionality.

\begin{itemize}
  \item \textbf{ACTi Camera} -- Saves still frames from ACTi IP surveillance cameras to disk at regular intervals.
  \item \textbf{ACU} -- The antenna control unit (ACU) agent. Controls the telescope platform pointing \cite{acu}.
  \item \textbf{Aggregator} -- Saves HK data from all other agents to \texttt{.g3} files on disk\footnote{\url{https://github.com/CMB-S4/spt3g_software}}.
  \item \textbf{Bluefors} -- Reads the Bluefors LD400 logs and passes them to \ocs{}.
  \item \textbf{Cryomech CPA} -- Monitors and controls the Cryomech helium compressors used with the pulse tube cryocoolers.
  \item \textbf{FTS Aerotech} -- Controls motion of a Fourier transform spectrometer (FTS) mirror stage.
  \item \textbf{Generator} -- Monitors the diesel generators used to power the site.
  \item \textbf{Hi6200} -- Records readings from an Hi6200 weight sensor. Used to monitor the amount of liquid nitrogen in cold traps on each platform.
  \item \textbf{HWP Encoder BBB} -- Receives and decodes encoder readings sent from a BeagleBone Black (BBB) connected to the half wave plate (HWP) hardware. The encoder signal is used to determine the HWP angles.
  \item \textbf{HWP Gripper} -- Monitors and controls three LEY32C-30 linear actuators. Used to `grip' the HWP when it is not floating on the superconducting bearing.
  \item \textbf{HWP PCU} -- Applies a discrete phase compensation to the HWP motor drive circuit.
  \item \textbf{HWP Picoscope} -- Monitors the position and temperature of the HWP with a Picoscope 3403D MSO connected to a set of LC sensors.
  \item \textbf{HWP PID} -- Interfaces with hardware that controls the rotation speed of the HWP with the use of a PID loop.
  \item \textbf{HWP PMX} -- Interfaces with a PMX Kikusui power supply, controlling the rotation of the HWP.
  \item \textbf{HWP Supervisor} -- Coordinates the rest of HWP agents. Can trigger shutdown of HWP system if needed.
  \item \textbf{Host Manager} -- Orchestrates the startup and shutdown of other OCS agents on a given node.
  \item \textbf{Holography FPGA} -- Interfaces with a ROACH2 FPGA to perform holography measurements in lab.
  \item \textbf{Holography Synthesizer} -- Interfaces with a frequency synthesizer to perform holography measurements in lab \cite{2021ApOpt..60.9029C}.
  \item \textbf{iBootbar} -- Controls the outlet state of network connected power distribution units (PDUs). Used to remotely reboot hardware when needed.
  \item \textbf{InfluxDB} -- Saves HK data from all other agents to the InfluxDB time series database for viewing in Grafana.
  \item \textbf{IFM SBN246 Flowmeter} -- Monitors the flow and temperature of liquid cooling loops used to cool hardware on each platform.
  \item \textbf{LabJack} -- Monitors analog and digital inputs and outputs of a LabJack T4 or T7. Used for miscellaneous sensors around the site.
  \item \textbf{Lakeshore 240} -- Interfaces with a Lakeshore 240 for monitoring the resistance and corresponding temperature of low temperature thermometry (both Ruthenium oxide and silicon diodes) at 1K and higher.
  \item \textbf{Lakeshore 336} -- Interfaces with a Lakeshore 336 for monitoring and servoing a cold load used for lab testing.
  \item \textbf{Lakeshore 370} -- Interfaces with a Lakeshore 370 for monitoring the resistance and corresponding temperature Ruthenium oxide thermometers down to 10 mK.
  \item \textbf{Lakeshore 372} -- Interfaces with a Lakeshore 372 for monitoring the resistance and corresponding temperature Ruthenium oxide thermometers down to 10 mK.
  \item \textbf{Lakeshore 425} -- Interfaces with a Lakeshore 425 Gaussmeter for measuring magnetic fields. Used to monitor the HWP float and rotation states.
  \item \textbf{LATRt XY Stage} -- Controls motors connected to an XY stage used for testing in the Large Aperture Telescope Receiver tester (LATRt) in lab.
  \item \textbf{Magpie} -- Receives detector timestreams from a SMuRF streamer and translates the streams into \texttt{G3Frames} that are compatible with the lyrebird detector monitor.
  \item \textbf{Meinberg M1000} -- Monitors the Meinberg LANTIME M1000, a GPS referenced clock that serves as the main source of timing for the entire site.
  \item \textbf{Meinberg Syncbox} -- Monitors the synchronization of the Meinberg Syncbox/N2X on each telescope platform. These receive a timing signal from the M1000 via PTP and translate it into a signal usable by other on platform hardware (IRIG, PPS, etc.)
  \item \textbf{Pfieffer TPG 366} -- Monitors a Pfieffer TPG 366, a six channel pressure gauge control unit used in the cryostats.
  \item \textbf{Pfieffer TC 400} -- Monitors and controls a Pfieffer TC 400 turbo pump controller.
  \item \textbf{Pysmurf Controller} -- Interfaces with the SMuRF crates to coordinate detector operations and readout.
  \item \textbf{Pysmurf Monitor} -- Monitors files created by the SMuRF system and tracks them for later transfer by the SupRsync agent.
  \item \textbf{Registry} -- Monitors the state of all other agents on the network as well the status of their operations. Provides a feed which can trigger alerts if critical operations fail.
  \item \textbf{SCPI PSU} -- Interfaces with any desktop power supply that is SCPI compatible, allowing for remote voltage and current control.
  \item \textbf{SMuRF Crate Monitor} -- Monitors the SMuRF Advanced Telecommunications Computing Architecture (ATCA) crates that power and connect the SMuRF blades, site networking, and timing infrastructure.
  \item \textbf{SMuRF Timing Card} -- Monitors the SMuRF timing card, which distributes timing information to all SMuRF systems on site.
  \item \textbf{SupRsync} -- Transfers files from the SMuRF servers to the RAID array on site for further aggregation using rsync.
  \item \textbf{Synaccess} -- Controls the outlet state of network connected PDUs. Used to remotely reboot hardware when needed.
  \item \textbf{Tektronix AWG} -- Controls a Tektronix AFG3021C function generator for use in lab. 
  \item \textbf{Thorlabs MC2000B} -- Monitors and controls a Thorlabs MC2000B optical chopper system.
  \item \textbf{UCSC Radiometer} -- Monitors the precipitable water vapor reported by a radiometer provided by The Universidad Cat\'olica de la Sant\'isima Concepci\'on (UCSC).
  \item \textbf{UPS} -- Monitors the uninterruptible power supplies (UPSes) supporting various on site hardware such as the computers.
  \item \textbf{Vantage Pro2} -- Monitors a Davis Instruments Vantage Pro2 weather monitor. Provides local site weather information such as temperature, wind speed, and UV index.
  \item \textbf{Wiregrid Actuator} -- Controls linear actuators used to insert and eject a wiregrid calibrator on the SATs.
  \item \textbf{Wiregrid Encoder} -- Monitors an encoder used to record the rotational angle of the wiregrid calibrator.
  \item \textbf{Wiregrid Kikusui} -- Monitors and controls a PMX Kikusui power supply that controls the rotation of the wiregrid calibrator.
  \item \textbf{Wiregrid Tilt Sensor\footnote{Not yet merged into \socs{} as of this writing, but nearly complete.}} -- Monitors a tilt sensor to record the tilt of the wiregrid calibrator against gravity.
\end{itemize}

\acknowledgments %
This work was supported in part by a grant from the Simons Foundation (Award
\#457687, B.K.). This work was supported by the National Science Foundation (UEI
GM1XX56LEP58). We would also like to thank the various open source projects
that we use to support observatory operations.

\bibliography{report} %
\bibliographystyle{spiebib} %

\end{document}